\documentclass[submission, Proceedings]{SciPost}
\usepackage[symbol]{footmisc}

\begin{document}

\begin{center}{\Large \textbf{
Solar Neutrino Physics with Borexino
}}\end{center}

\begin{center}
A.~Pocar\textsuperscript{1,\footnote[1]{}}
M.~Agostini\textsuperscript{2}
K.~Altenm\"{u}ller\textsuperscript{2}
S.~Appel\textsuperscript{2}
V.~Atroshchenko\textsuperscript{3}
Z.~Bagdasarian\textsuperscript{4}
D.~Basilico\textsuperscript{5}
G.~Bellini\textsuperscript{5}
J.~Benziger\textsuperscript{6}
G.~Bonfini\textsuperscript{7}
% ...
\end{center}

\begin{center}
{\bf 1} Amherst Center for Fundamental Interactions and Physics Department, University of Massachusetts, Amherst, MA 01003, USA
\\
{\bf 2} Physik-Department and Excellence Cluster Universe, Technische Universit\"at  M\"unchen, 85748 Garching, Germany
\\
{\bf 3} National Research Centre Kurchatov Institute, 123182 Moscow, Russia
\\
{\bf 4} Institut f\"ur Kernphysik, Forschungszentrum J\"ulich, 52425 J\"ulich, Germany
\\
{\bf 5} Dipartimento di Fisica, Universit\`a degli Studi e INFN, 20133 Milano, Italy
\\
{\bf 6} Chemical Engineering Department, Princeton University, Princeton, NJ 08544, USA
\\
{\bf 7} INFN Laboratori Nazionali del Gran Sasso, 67010 Assergi (AQ), Italy
\\
{\bf 8} Physics Department, Princeton University, Princeton, NJ 08544, USA
\\
{\bf 9} Dipartimento di Fisica, Universit\`a degli Studi e INFN, 16146 Genova, Italy
\\
{\bf 10} Physics Department, Virginia Polytechnic Institute and State University, Blacksburg, VA 24061, USA
\\
{\bf 11} Lomonosov Moscow State University Skobeltsyn Institute of Nuclear Physics, 119234 Moscow, Russia
\\
{\bf 12} Department of Physics and Astronomy, University of Hawaii, Honolulu, HI 96822, USA
\\
{\bf 13} St. Petersburg Nuclear Physics Institute NRC Kurchatov Institute, 188350 Gatchina, Russia
\\
{\bf 14} Gran Sasso Science Institute, 67100 L'Aquila, Italy
\\
{\bf 15} Joint Institute for Nuclear Research, 141980 Dubna, Russia
\\
{\bf 16} AstroParticule et Cosmologie, Universit\'e Paris Diderot, CNRS/IN2P3, CEA/IRFU, Observatoire de Paris, Sorbonne Paris Cit\'e, 75205 Paris Cedex 13, France
\\
{\bf 17} Kepler Center for Astro and Particle Physics, Universit\"{a}t T\"{u}bingen, 72076 T\"{u}bingen, Germany
\\
{\bf 18} Department of Physics, University of Houston, Houston, TX 77204, USA
\\
{\bf 19} M.~Smoluchowski Institute of Physics, Jagiellonian University, 30348 Krakow, Poland
\\
{\bf 20} RWTH Aachen University, 52062 Aachen, Germany
\\
{\bf 21} Kiev Institute for Nuclear Research, 03680 Kiev, Ukraine
\\
{\bf 22} National Research Nuclear University MEPhI (Moscow Engineering Physics Institute), 115409 Moscow, Russia
\\
{\bf 23} Institute of Physics and Excellence Cluster PRISMA, Johannes Gutenberg-Universit\"at Mainz, 55099 Mainz, Germany
\\
{\bf 24} Department of Physics, Technische Universit\"at Dresden, 01062 Dresden, Germany
\\
{\bf 25} Dipartimento di Chimica, Biologia e Biotecnologie, Universit\`a degli Studi e INFN, 06123 Perugia, Italy
\\
\footnotemark[1]{ Corresponding Author, pocar@umass.edu}
\\
\footnotemark[2]{Present address: Universidad Autonoma de Madrid, Ciudad Universitaria de Cantoblanco, 28049 Madrid, Spain}
\\
\footnotemark[3]{Present address: INFN Laboratori Nazionali del Gran Sasso, 67010 Assergi (AQ), Italy}
\\
\footnotemark[4]{Present address: Fermilab National Accelerato Laboratory (FNAL), Batavia, IL 60510, USA}
\\
\footnotemark[5]{Present address: Dipartimento di Fisica, Sapienza Universit\`a di Roma e INFN, 00185 Roma, Italy}
\\
\footnotemark[6]{Present address: Dipartimento di Fisica, Universit\`a degli Studi Federico II e INFN, 80126 Napoli, Italy}
\end{center}

\begin{center}
\today
\end{center}


\section*{Abstract}
{\bf
We present the most recent solar neutrino results from the Borexino experiment at the Gran Sasso underground laboratory. In particular, refined measurements of all neutrinos produced in the {\it pp} fusion chain have been made. It is the first time that the same detector measures the entire range of solar neutrinos at once. These new data weakly favor a high-metallicity Sun. Prospects for measuring CNO solar neutrinos are also discussed.
}

\vspace{10pt}
\noindent\rule{\textwidth}{1pt}
\tableofcontents\thispagestyle{fancy}
\noindent\rule{\textwidth}{1pt}
\vspace{10pt}

\section{Introduction: solar neutrinos and Borexino}
\label{sec:intro}
\begin{figure}[!h]
\begin{center}
\includegraphics[height=2.5in]{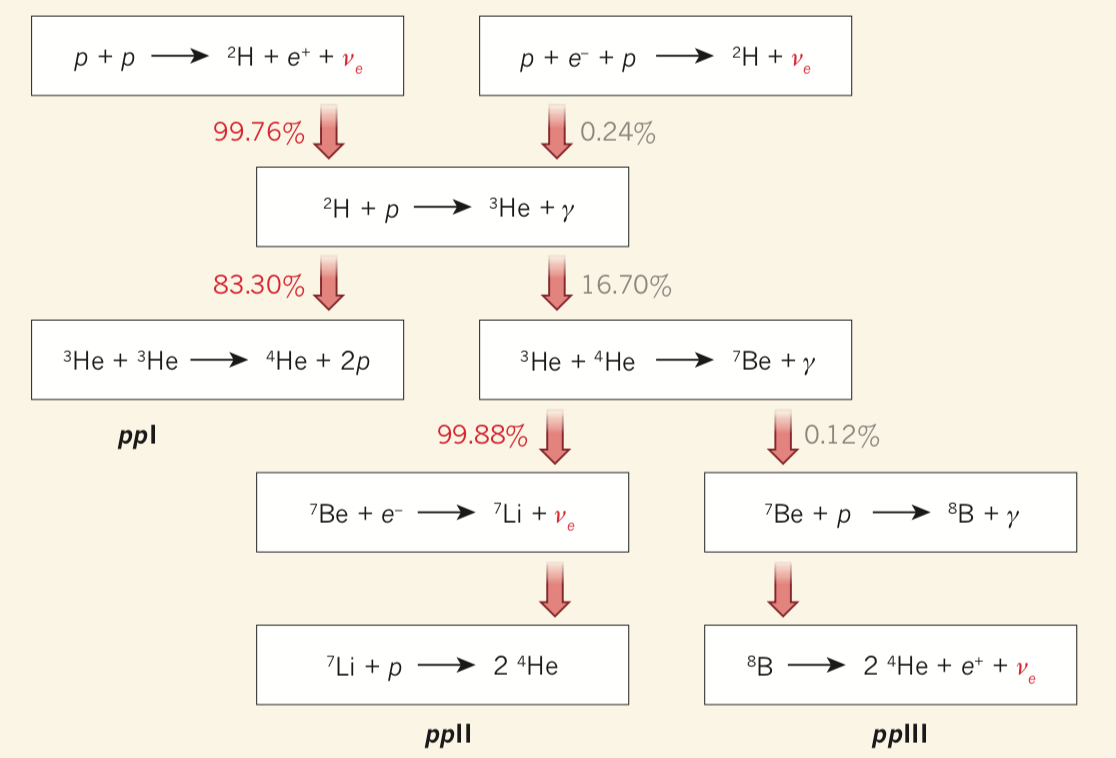}
\end{center}
\vspace{-0.4cm}\caption{The solar {\it pp} fusion chain.} 
\label{f:pp}
\end{figure}

\begin{figure}[!b]
\begin{center}
\includegraphics[height=1.6in]{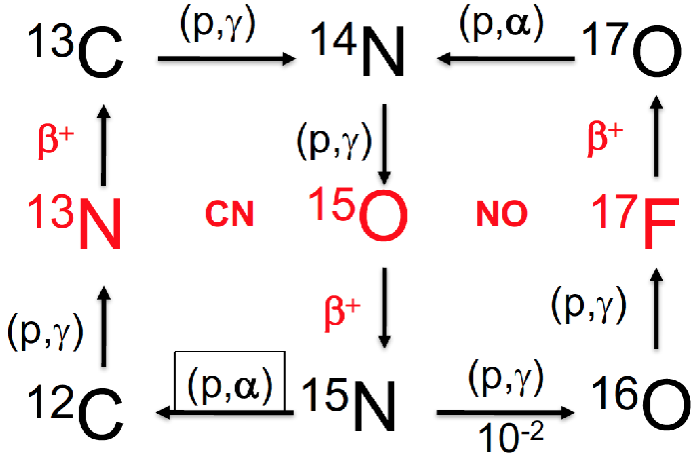}
\hspace{0.01in}
\includegraphics[height=1.7in]{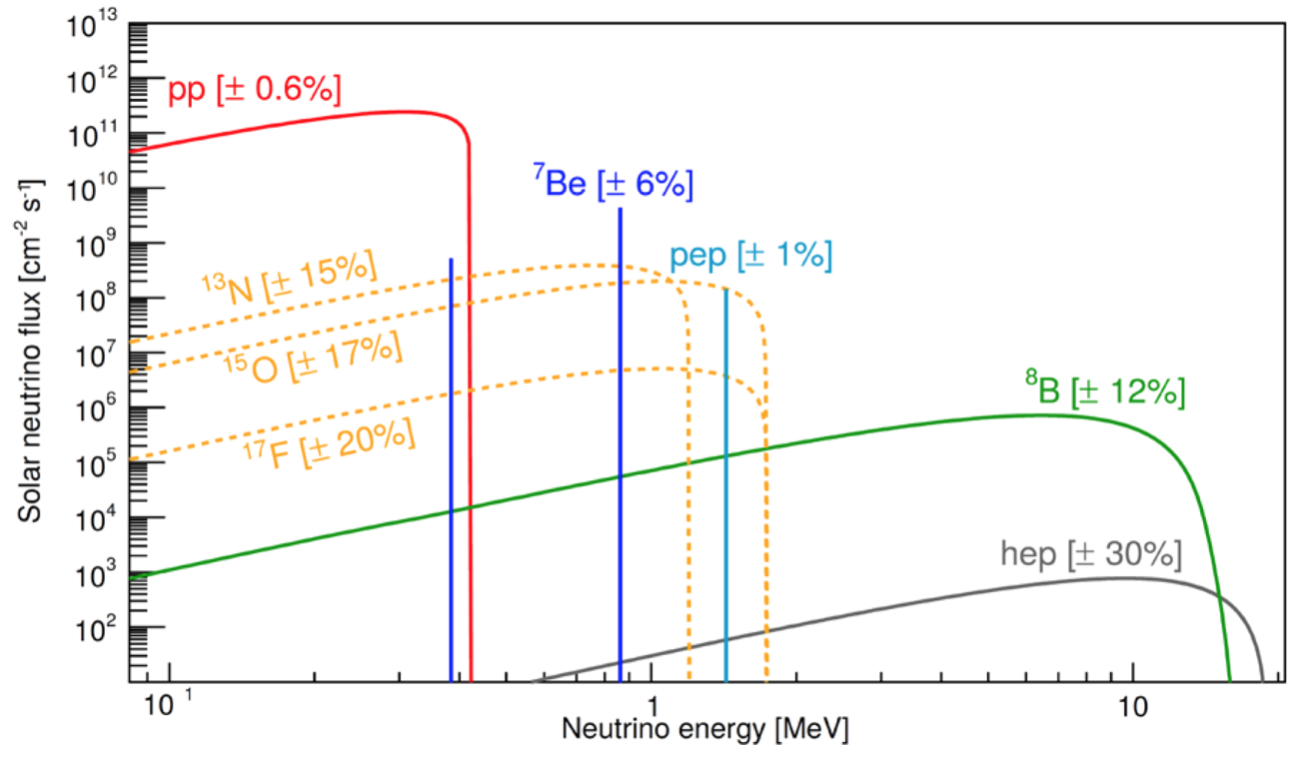}
\end{center}
\vspace{-0.4cm}\caption{Left: the CNO solar fusion cycle. Right:the SSM solar neutrino spectrum.} 
\label{f:cno-spectrum}
\end{figure}

We know that the Sun is fueled by nuclear reactions fusing protons (hydrogen) into helium. 
The set of reactions accomplishing this is summarized as:
$$
4p\,\longrightarrow\,^4He+2e^+ +2\nu_e + (24.7 + 2m_ec^2)[MeV].
$$
In the Sun, 99\% of the times this process is carried out through a set of reactions known as the {\it pp}-chain, initiated by the fusion of two protons as illustrated in Fig.~\ref{f:pp}. 
Fig.~\ref{f:cno-spectrum} shows the reactions believed to contribute the remaining $\sim$1\%, in which proton fusion is catalized by heavier elements, enhanced by higher {\it metallicity} (in astrophysics, all elements heavier than helium are call {\it metals}). 
Also shown is the spectrum of solar neutrinos predicted by the Standard Solar Model (SSM).
A comprehensive review of solar neutrino physics, with connections to their experimental investigation, their role in the discovery of neutrino oscillations, and the definition of neutrino flavor conversion parameters is found in~\cite{Haxton:2013cx}.

\begin{figure}[!t]
\begin{center}
\includegraphics[height=3.5in]{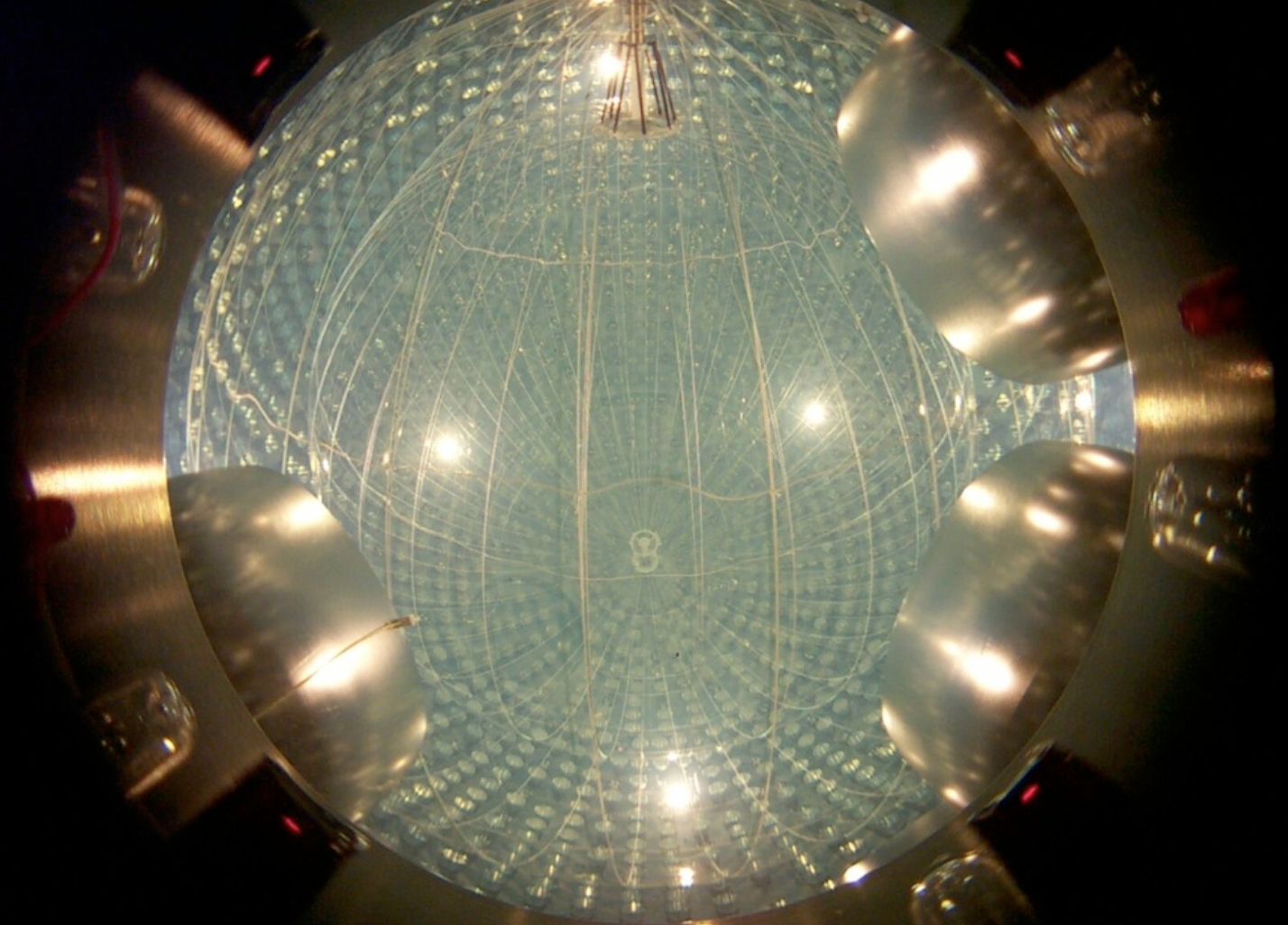}
\end{center}
\vspace{-0.4cm}\caption{The Borexino inner detector filled with scintillator in 2007.} 
\label{f:bx}
\end{figure}

Borexino measures solar (and other low energy) neutrinos interacting with a spherical target of $\sim$300 tonnes of organic liquid scintillator. 
The scintillator is contained within a thin, transparent nylon vessel and is surrounded by $\sim$1 kilo-tonne of buffer liquid. 
Scintillation pulses from neutrino interactions, as well as other (mostly background) ionizing events are detected by $\sim$2,000 8-inch photomultiplier tubes uniformly mounted to point radially inwards on a 13-meter diameter stainless steel sphere containing the fluid buffer. 
A water tank operated as a \v{C}erenkov muon detector hermetically surrounds the inner detector.
Neutrinos are detected via their elastic scattering off the electrons of the scintillator. 
The use of liquid scintillator allows one to operate with a very low threshold compared to other techniques. 
The scintillation signal, however, is isotropic and does not thus preserve directional information of events.
For this reason ({\it i.e.} the lack of specificity of neutrino interactions with respect to that of most backgrounds), the detector requires unprecedented purity from any radioactivity in the scintillator and all construction materials.
A picture of the Borexino scintillator target and inner detector is shown in Fig.~\ref{f:bx}.
The Borexino detector and its design characteristics are described in detail in~\cite{Alimonti:2009vb}.

In the following, we provide a summary of the main results from Borexino. Sec.~\ref{s:history} gives an overview of pre-2018 results, Sec.~\ref{s:recent} reports on recent measurements that cover the entire {\it pp} chain, Sec.~\ref{s:cno} provides a status update on attempting to measure CNO solar neutrinos, and Sec.~\ref{s:conclusions} offers brief conclusions.

\section{Brief summary of Borexino results}
\label{s:history}
Borexino has been running continuously since May 2007, measuring the entire solar neutrino energy spectrum with increasing precision. 
It immediately showed unprecedented low levels of radioactive background (in the form of $^{40}$K, $^{232}$Th, $^{238}$U) which made it possible to observe the existence of $^7$Be solar neutrinos with just a few weeks of data. 
Over the following three years (the so-called Phase I), Borexino has:
\begin{itemize}
\item measured the interaction rate of $^7$Be neutrinos with better than 5\% precision, {\it i.e.} more precise than the SSM uncertainty;
\item excluded, with high sensitivity, any day-night asymmetry in the $^7$Be solar neutrino flux, setting a stringent limit on electron flavor regeneration for solar neutrinos traversing the earth and providing further confirmation of the validity of the MSW-LMA solar neutrino oscillation model;
\item provided evidence on the existence of {\it pep} neutrinos;
\item yielded the strongest upper bound for the CNO solar neutrino interaction rate;
\item measured $^8$B solar neutrinos with the lowest detection energy threshold;
\item tested the existence of a variety of ultra-rare, non-standard processes such as the existence of a neutrino magnetic moment.
\end{itemize}
These results are included for reference in the table in Fig~\ref{f:results}. 
An account of the Borexino Phase-I results and analysis methods can be found in Ref.~\cite{Bellini:2014un} and references therein. \\

Following a period of about two two years during which the Borexino scintillator was further purified from radioactivity, in 2012 the detector began its Phase II data collection. 
The purification achieved the goal of reducing $^{232}$Th, $^{238}$U contamination to the $10^{-19}$ gram-of-contaminant per gram-of-scintillator level, three orders of magnitude lower than the initial Borexino specifications.
Noteworthy results prior to 2017 include:
\begin{itemize}
\item the first spectral measurement of {\it pp} solar neutrinos~\cite{Collaboration:2014th};
\item evidence that the $^7$Be neutrino interaction rate displayed a seasonal modulation consistent with the varying solid angle between the earth and the sun\cite{Agostini:2017bk};
\item the detection of geo-neutrinos ({\it i.e.} anti-neutrinos from radioactivity in the earth) with high significance, along with the $\sim$1,000 km-baseline detection of anti-neutrinos from distant nuclear reactors in France~\cite{Borexino:2015un};
\item searches for neutrino signals in coincidence with gamma ray bursts and gravitational waves~\cite{Agostini:2016hb,Agostini:2017ji}.
\end{itemize}

\section{Latest results from Borexino}
\label{s:recent}
Phase II Borexino data provided the opportunity to perform higher precision measurements of all solar neutrino fluxes~\cite{BorexinoCollaboration:2018bt}. 
Firstly, reduced radioactive backgrounds with respect to Phase I, specifically $^{85}$Kr, $^{210}$Bi, and $^{210}$Po contributed to a better measurement of $^7$Be neutrinos. 
Radio-purity, however, was not the only factor in better measurements.
A greatly-improved Monte Carlo simulation package was developed~\cite{Agostini:2018bu} that allowed a more accurate determination of the energy response over a wide energy range, and background-suppressing analysis tools were refined, adding to the intrinsic improvement provided by an extended data set.

\begin{figure}[!h]
\begin{center}
\includegraphics[height=3in]{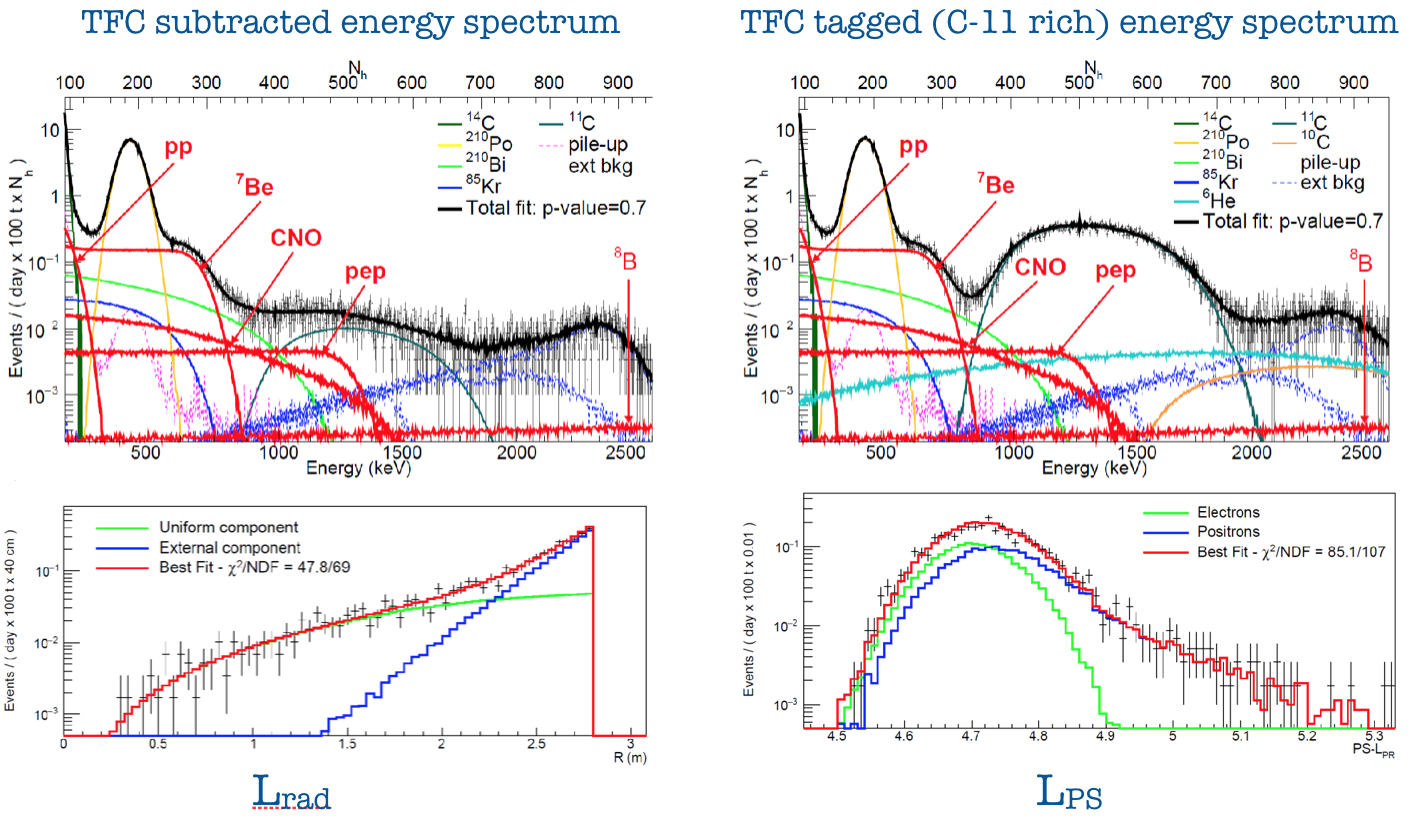}
\end{center}
\vspace{-0.4cm}\caption{Illustration of the Phase II Borexino fitting strategy used to simultaneously determine the interaction rate of {\it pp}, $^7$Be, and {\it pep} solar neutrinos. The fit also returns an upper limit for CNO neutrinos, as well as rates for all relevant background sources.} 
\label{f:fit}
\end{figure}

A multi-variate approach was used to identify and suppress the cosmogenic $^{11}$C background ($\sim$30 minute half life)), a $\beta^+$ emitter covering the energy range relevant for {\it pep} and CNO neutrino detection.
Positron emission with the production (50\% of the times) of 3 ns-lived ortho-positronium and the production of annihilation gamma rays (extended and with a slightly different ionization density profile in the scintillator) produces a statistically distinguishable time profile of the scintillation pulse from that of electron events.
A likelihood was built using such a pulse shape parameter, the radial distribution of events, and the simultaneous fit of $^{11}$C-rich and $^{11}$C-subtracted energy spectra.
This procedure for determining neutrino fluxes is illustrated in Fig.\ref{f:fit}, and allowed us to simultaneously fit the Borexino data between $\sim$200 keV and $\sim$2.5 MeV, including the interaction rate of {\it pp}, $^7$Be, {\it pep}, and CNO solar neutrinos and the overlapping backgrounds (previous measurements were carried out focusing on narrower energy regions).

The $^8$B solar neutrino rate was measured separately, albeit with the same data set. 
This choice was dictated by the need to boost the statistics for this dimmer solar neutrino component.
In fact, the entire scintillator volume was used, instead of an innermost fiducial volume.
This choice was possible due to the higher energy of most of the $^8$B spectrum, placing a lower energy cut of 3.2 MeV, above much of the natural radioactivity. \\

\begin{figure}[!t]
\begin{center}
\includegraphics[height=2.2in]{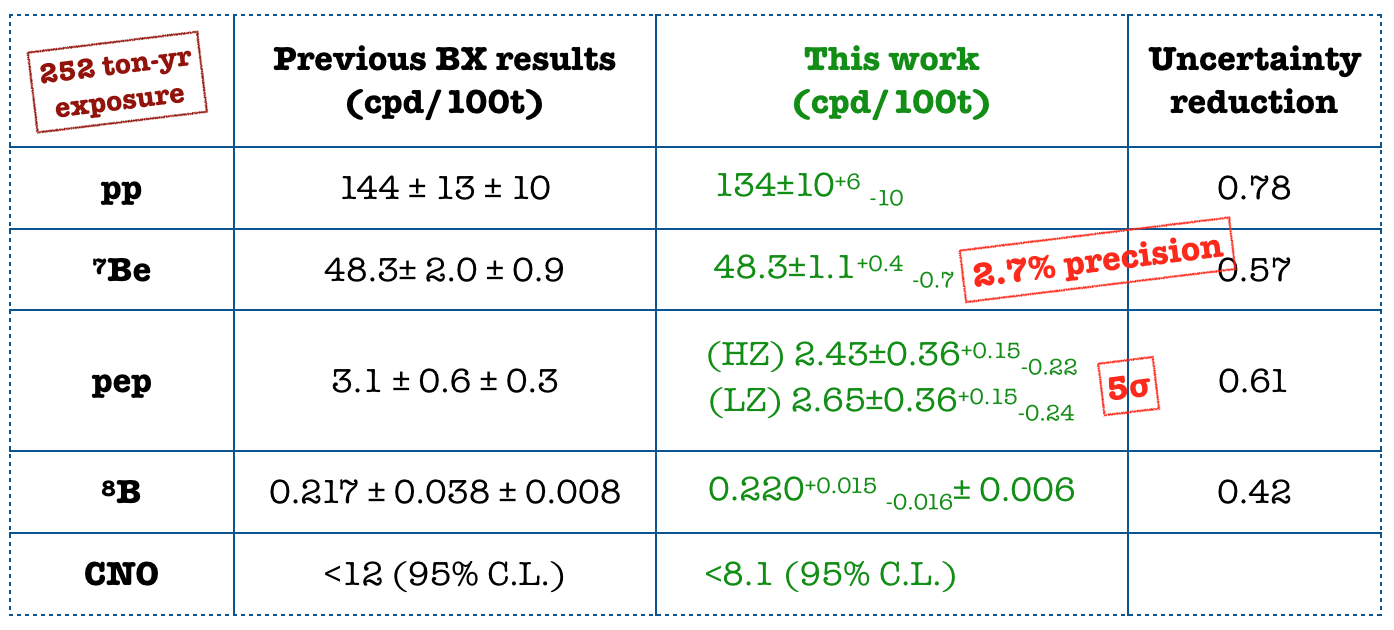}
\end{center}
\vspace{-0.4cm}\caption{Phase-II Borexino solar neutrino results compared to pre-2018 measurements.} 
\label{f:results}
\end{figure}

The Phase II results are collected in the table in Fig.~\ref{f:results}. 
The $^7$Be neutrino interaction rate is now measured with a precision of $<$3\%.
This is twice as small than the theoretical uncertainty from the SSM, and can be used, in combination with assumptions on the solar neutrino oscillation parameters, to better our understanding of the sun.
The uncertainty on the {\it pp} neutrino rate is reduced to $\sim$10\%.
This is no small feat, given that its very measurement went beyond what Borexino had proposed to do and that it was believed a dedicated experiment was necessary to measure this low energy component of the flux. 
The {\it pep} neutrinos are definitively discovered with 5$\sigma$ significance, and a tight limit of $<8.1$ counts per day (cpd)/100t (95\% C.L.) is set for the CNO neutrino rate.
The $^8$B rate is not as precise as that measured by SuperK~\cite{Abe:2016el}, but is measured with the lowest energy threshold of all experiments for this component.\\

The current goal of solar neutrino physics is to use neutrinos to increase our understanding of solar models. 
Borexino Phase II data allows us for the first time to study the sun's metallicity, {\it i.e} the abundance of elements heavier than helium.
High- and low-metallicity solar models (referred to as HZ and LZ, respectively) result from contrasting measurements. 
Helio-seismological data prefer HZ solar photospheric abundances. 
Solar metallicity affects solar neutrino fluxes, most prominently that of CNO neutrinos with a $\sim$30\% higher flux predicted by the HZ model compared to the LZ one.
However, both the $^7$Be and $^8$B fluxes are $\sim$10\% higher in the HZ SSM, while the {\it pp} and {\it pep} fluxes are higher for the LZ SSM~\cite{Haxton:2013cx}.
Thus the solar metallicity question could be addressed with high-precision measurements of these fluxes.

The $^7$Be and {\it pp} neutrino fluxes can be used to compare the relative weight of the two helium-helium fusion reactions by computing 
$$
R \equiv \frac{^3 {\rm He}+ ^4{\rm He}}{^3{\rm He} + ^3{\rm He}} = \frac{2\phi(^7{\rm Be})}{\phi(pp)+\phi(^7{\rm Be})},
$$
which is predicted to be ($0.180\pm0.11$) and  ($0.161\pm0.10$) for HZ and LZ SSM, respectively.
This ratio measured by Borexino is $R_{BX}=0.178^{+0.027}_{-0.023}$.

\begin{figure}[!h]
\begin{center}
\includegraphics[height=1.8in]{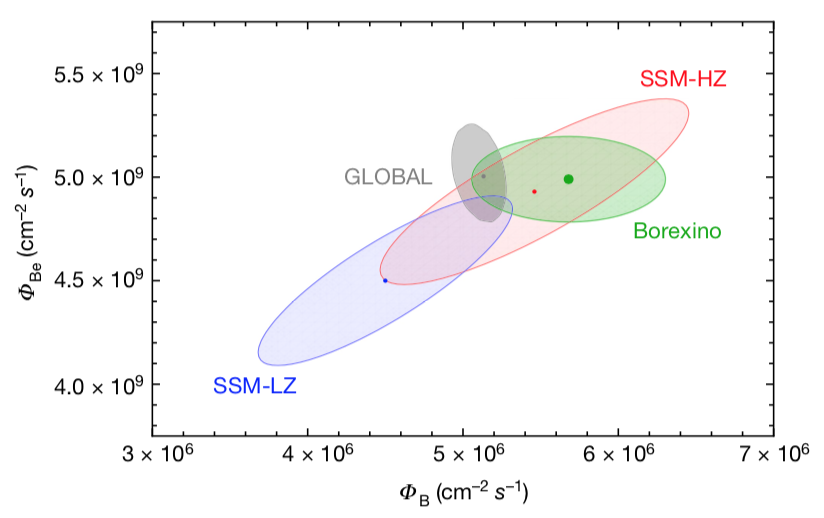}
\hspace{0.01in}
\includegraphics[height=1.7in]{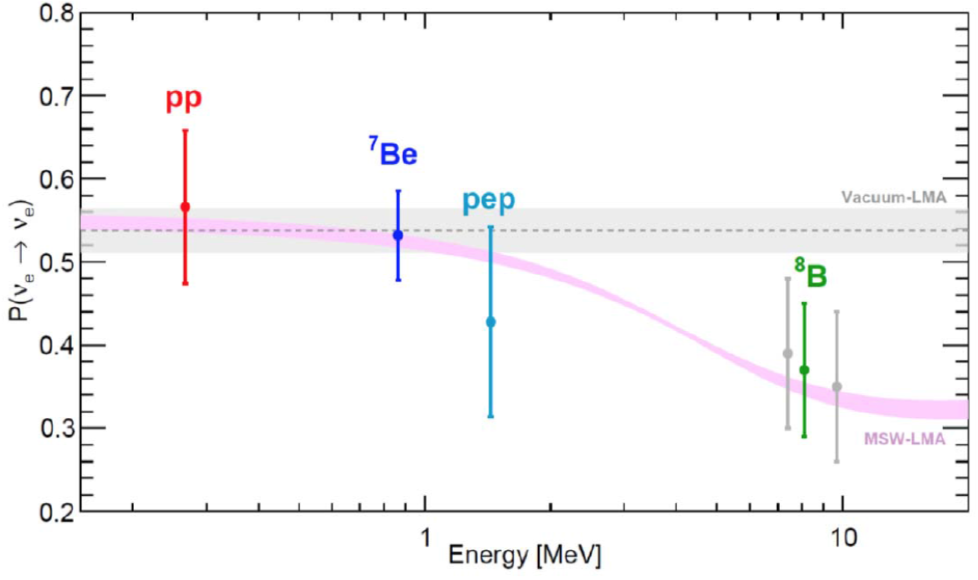}
\end{center}
\vspace{-0.4cm}\caption{Left: Phase-II Borexino and solar neutrino experiments $^7$Be and $^8$B solar neutrino results compared against theoretical predictions from HZ and LZ SSM. Ellipses indicate 1 $\sigma$ contours. Right: the electron flavor survival probability of the solar neutrino components measured with Borexino. The pink band represents the energy-dependent survival probability under the assumption of HZ SSM.}
\label{f:ellipses}
\end{figure}

Similarly, the measured $^7$Be and $^8$B can be used together and compared with SSM HZ and LZ cases, as shown in Fig.~\ref{f:ellipses}. 
Borexino data alone mildly prefer the HZ SSM. 
This hint is weakened by including all solar neutrino data, which notably provides a more precise value for the $^8$B neutrinos as measured by the SuperK experiment, in the analysis.
In addition, theoretical uncertainties barely differentiate the two scenarios in this case.

\section{Outlook: measuring CNO neutrinos}
\label{s:cno}
Of great astrophysical interest is the measurement of CNO neutrinos, as they are arguably the most direct probe of solar metallicity.
In order to assess how far Borexino is from measuring CNO neutrinos, the current Borexino limit $<8.1$ cpd/100t (95\% C.L.) should be compared with SSM HZ and LZ predictions of  $4.91\pm0.52$ cpd/100t and $3.52\pm0.37$ cpd/100t, respectively.
On one hand, this extra factor of two seems at arm's reach of the experiment. 
On the other, one background exists that poses a serious challenge, the $\beta$ emitter $^{210}$Bi.

\begin{figure}[!h]
\begin{center}
\includegraphics[height=1.7in]{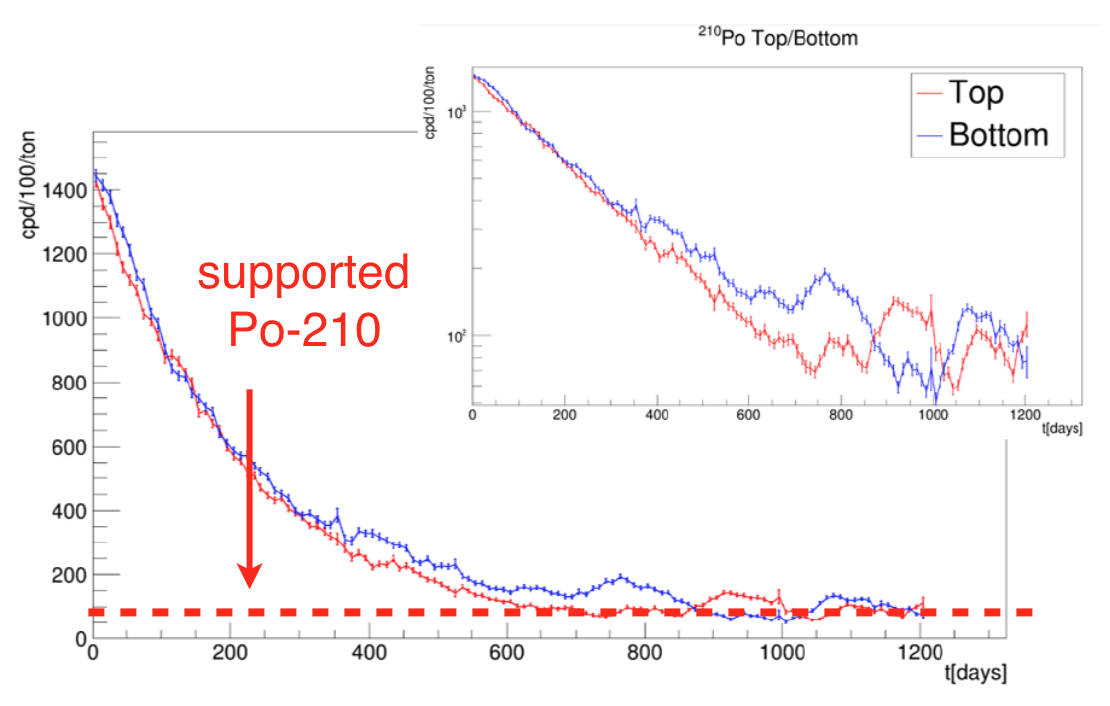}
\hspace{0.01in}
\includegraphics[height=1.7in]{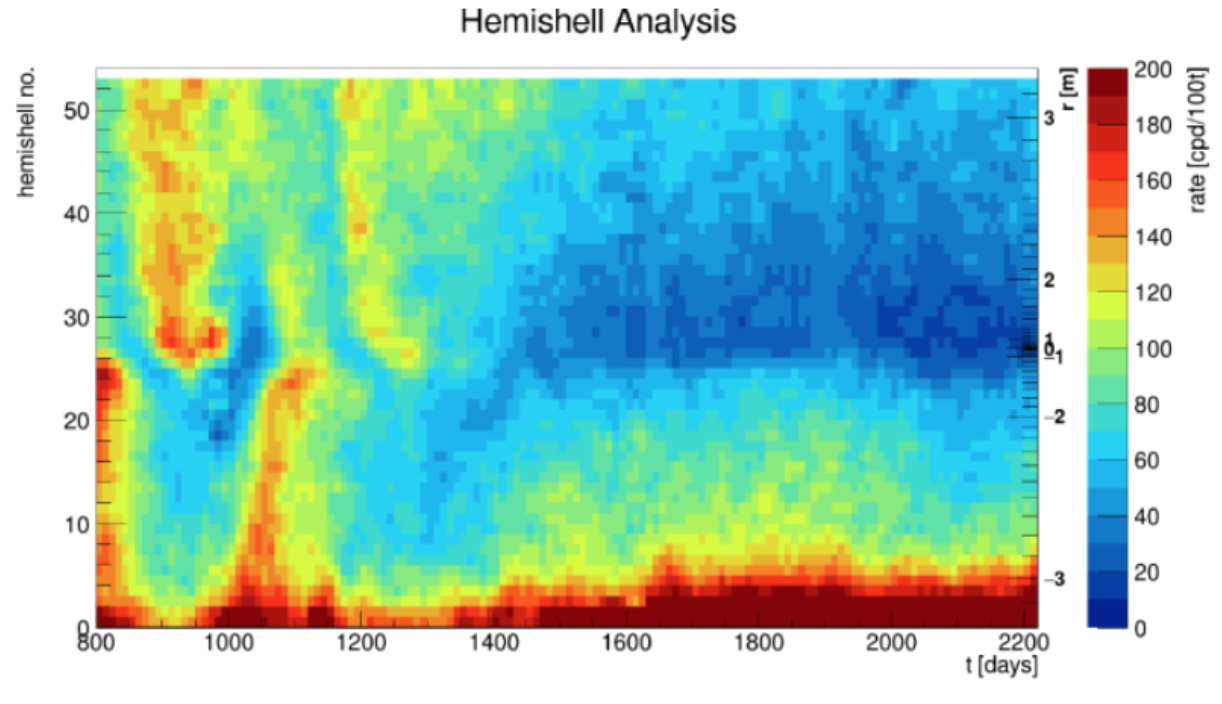}
\end{center}
\vspace{-0.4cm}\caption{Left: The $^{210}$Po activity in the inner part of the Borexino scintillator volume versus time for the Phase II dataset. The data is divided between top and bottom half of the volume. It shows deviations from a pure exponential decay with the $^{210}$Po half life (138 days), a behavior consistent with 'dirtier' scintillator from outside the fiducial volume entering and leaving it as a result of convective motions. Right: the effectiveness of thermally insulating the Borexino tank is evident from the ensuing layering of the $^{210}$Po activity inside the fiducial volume.} 
\label{f:po210}
\end{figure}

$^{210}$Bi is a decay product of $^{222}$Rn, and because of this is found in air and on virtually all surfaces. 
It is sustained by its long-lived daughter $^{210}$Pb, and is followed by a relatively long-lived $\alpha$ emitter, $^{210}$Po.
The $^{210}$Bi $\beta$ spectrum is quasi-degenerate with that of electrons recoiling off CNO neutrinos, as shown in Fig.~\ref{f:fit}.
Detection of CNO neutrinos thus hinges on having very low and well measured $^{210}$Bi background. 
One could determine the $^{210}$Bi activity by measuring the supported $^{210}$Po component after the fraction which is out of equilibrium has decayed away, as proposed in Ref.~\cite{Villante:2011wj}.
Fig.~\ref{f:po210} shows the $^{210}$Po activity in the Borexino fiducial volume versus time for approximately the past three years.
A precise determination of the steady-state component is made difficult by background fluctuations caused by scintillator mixing due to convective motions with timescales of several months.

In 2015, the entire Borexino detector was thermally insulated from the air of the experimental hall, the effect of which can be appreciated in Fig.~\ref{f:po210}.
The collaboration is looking hard into whether, with this stabler detector, $^{210}$Bi is low and constrained enough to allow for a measurement of CNO neutrinos in the near future.

\section{Conclusion}
\label{s:conclusions}
Borexino has been running for more than ten years, measuring the entire solar neutrino energy spectrum with increasing precision. Noteworthy is the first measurement of $^7$Be neutrinos, the prominent piece of the "solar neutrino puzzle", eventually solved by the discovery of neutrino flavor transformation. 

Borexino has set a new standard in levels of trace levels of radioactive impurities for large, rare event experiments, making it possible to produce the crispest measurement to date of solar neutrinos over their entire spectrum.
The precision on the $^7$Be rate is a factor of two better that the SSM uncertainty on this component. 
The improved measurement of the {\it pp} and {\it pep} rates, which are consistent with each other based on known nuclear physics cross sections, and consistent with our understanding of the sun and the solar luminosity, reinforce our ability to experimentally confirm that the sun is in thermodynamic equilibrium on a scale of $10^4 - 10^5$ years\cite{Haxton:2014um}. 

In the next few years, Borexino will attempt to measure the CNO solar neutrino interaction rate, which could resolve the solar metallicity puzzle, an important open question about the sun.

\section*{Acknowledgements} 
The Borexino Collaboration acknowledges the generous hospitality and support of the Laboratori Nazionali del Gran Sasso (Italy).


\paragraph{Funding information}
The Borexino program is made possible by funding from INFN (Italy), NSF (USA), BMBF, DFG (OB168/2-1, WU742/4-1, ZU123/18-1), HGF, and MPG (Germany), RFBR (Grants 16-02-01026 A, 15-02-02117 A, 16-29-13014 ofim, 17-02-00305 A), RSF (Grant 17-02-01009) (Russia), and NCN (Grant UMO 2013/10/E/ST2/00180) (Poland).

\bibliography{tau2018-bx-proceedings-pocar}

\nolinenumbers

\end{document}